# Probing New Physics from the Single-Top Production[a]

Douglas O. Carlson and C.–P. Yuan

*Department of Physics and Astronomy*
*East Lansing, MI 48824 , USA*

We show that the production rate of the single-top quark event from the $W$-gluon fusion process is sensitive to new physics which strongly modifies the interactions of the top quark. The measurement of its production rate gives the partial decay width of the top quark which combined with the branching ratio measurement from the $t\bar{t}$ pair event determines the lifetime of the top quark.

## 1  Introduction

From the direct search at the Tevatron, the top quark has been discovered and found to have mass of $m_t = 176 \pm 8\,(\text{stat.}) \pm 10(\text{sys.})\,\text{GeV}$ from CDF data, and $m_t = 199^{+29}_{-21}\,(\text{stat.}) \pm 22(\text{sys.})\,\text{GeV}$ from DØ data. For a heavy top quark, $m_t$ is of the order of the electroweak symmetry breaking scale $v = \left(\sqrt{2}G_F\right)^{-1/2} = 246\,\text{GeV}$. ($v/\sqrt{2} = 175\,\text{GeV}$.) It is likely that the generation of fermion mass is closely related to the electroweak symmetry breaking, effects from new physics would be more apparent in the top quark sector than any other light sector of the electroweak theory. Thus, the top quark system may be used to probe the symmetry breaking sector. A few examples are discussed in Ref. [1] to illustrate that different models of electroweak symmetry breaking mechanism will induce different interactions among the top quark and the $W$- and $Z$-bosons. These interactions may strongly modify the production and/or the decay of the top quark. It is therefore important to test whether it is a Standard Model (SM) top quark.

In this work we will concentrate on the single-top quark event produced from the $W$-gluon fusion processes $q'g(W^+g) \to qt\bar{b}$ or $q'b \to qt$.

---

[a] Talk given by C.–P. Yuan at the Workshop on Physics of the Top Quark, IITAP, Iowa State University, Ames, Iowa, May 25 & 26, 1995.

Table 1: Cross sections (in pb) of various processes for $m_t = 180(140)$ GeV. (Branching ratios are not included.) For the single-top rates we only include single-$t$ production.

| $\sqrt{S}$(TeV) | $q\bar{q}, gg \to t\bar{t}$ | $W$-gluon fusion | $q'\bar{q} \to W^* \to t\bar{b}$ | $gb \to W^- t$ |
|---|---|---|---|---|
| 2 ($\bar{p}p$) | 4.5(16) | 1(2) | 0.3(0.8) | 0.1(0.3) |
| 14 (pp) | 430(1300) | 100(140) | 4.6(11) | 3.6(8.8) |

## 2   The Single-Top Production Mechanism

In this section we consider the production rate of a single-top quark at hadron colliders. In referring to single-top production, unless stated otherwise, we will concentrate only on the positive charge mode (*i.e.* only including single-$t$, but not single-$\bar{t}$). The colliders we consider are the Tevatron (a $\bar{p}p$ collider) with the Main Injector (Run-II) at $\sqrt{S} = 2$ TeV and the LHC (a pp collider) at $\sqrt{S} = 14$ TeV.

A single-top quark signal can be produced from the $W$-gluon fusion process $q'g(W^+g) \to qt\bar{b}$ (or $q'b \to qt$) [2], the Drell-Yan type process $q'\bar{q} \to W^* \to t\bar{b}$ (also known as "$W^*$" production) [3], or $Wt$ production via $gb \to W^- t$ [4]. For later reference in this paper, we show the total cross sections of these processes at the Tevatron and the LHC in Table 1, for $m_t = 180(140)$ GeV. For the single-top rates we only include $t$ production. To include the production rates for both single-$t$ and single-$\bar{t}$ events at $\bar{p}p$ colliders, a factor of 2 should be multiplied to the single-$t$ production rates because the parton luminosity for single-$\bar{t}$ production is the same as that for single-$t$. At the LHC the rates should be multiplied by $\sim 2$ because the relevant parton luminosities for producing single-$t$ and single-$\bar{t}$ events in high energy collisions are dominated by the sea quarks.

The single-top quark produced from the $W$-gluon fusion process involves a very important and not yet well-developed technique of handling the kinematics of a *heavy b* parton inside a hadron. Thus the kinematics of the top quark produced from this process cannot yet be accurately calculated. However, the total event rate for single-top quark production via this process can be estimated using the method proposed in Ref. [5]. The total rate for the $W$-gluon fusion process involves the $\mathcal{O}(\alpha^2)$ $(2 \to 2)$ process $q'b \to qt$ plus the $\mathcal{O}(\alpha^2\alpha_s)$ $(2 \to 3)$ process $q'g(W^+g) \to qt\bar{b}$ (where the gluon splits to $b\bar{b}$) minus the *splitting* piece $g \to b\bar{b} \otimes q'b \to qt$ in which $b\bar{b}$ are nearly collinear. These processes are shown diagrammatically in Fig. 1. The splitting piece is subtracted to avoid double counting the regime in which the $b$ propagator in the $(2 \to 3)$ process is close to on-shell. This procedure is to resum the large logarithm $\alpha_s \ln(m_t^2/m_b^2)$ in the $W$-gluon fusion process to all orders in $\alpha_s$ and include part of the higher order $\mathcal{O}(\alpha^2 \alpha_s)$ corrections to its production rate. ($m_b$ is the mass of the bottom quark.) We note that to obtain the complete $\mathcal{O}(\alpha^2 \alpha_s)$ corrections beyond just the leading log contributions one should also include virtual corrections to the $(2 \to 2)$ process, but we shall ignore these non-leading contributions here.[b] Using the prescription described as above we found that the total rate of the $W$-gluon fusion process is about 25% less than that from the $(2 \to 2)$ process for $m_t = 180\,(140)$ GeV regardless of the energy or the type (*i.e.* pp or p̄p) of the machine. To estimate the uncertainty in the production rate due to the choice of the scale $Q$ in evaluating the strong coupling constant $\alpha_s$ and the parton distribution function (PDF), we have performed a detailed study in Ref. [7]. Although the individual rate from either $(2 \to 2)$, $(2 \to 3)$ or the splitting piece is relatively sensitive to the choice of the scale, the total rate as defined by $(2 \to 2) + (2 \to 3) -$ (splitting piece) only varies by about 30% and 10% for $M_W/2 < Q < 2m_t$ at the Tevatron and the LHC, respectively. We note that in the $(2 \to 2)$ process the $b$ quark distribution effectively contains sums to order $[\alpha_s \ln(Q/m_b)]^n$ from $n$-fold collinear gluon emission, whereas the subtraction term (namely, the splitting piece) contains only first order in $\alpha_s \ln(Q/m_b)$. As $Q \to m_b$, the $(2 \to 2)$ process picks up only the leading order in $\alpha_s \ln(Q/m_b)$ and so is largely cancelled by the splitting piece in calculating the total rate. Consequently, the total rate is about the same as the $(2 \to 3)$ rate for $Q \to m_b$. We also note that when $Q \simeq M_W/2$, the $(2 \to 2)$ and $(2 \to 3)$ processes have about the same rate. As $Q$ increases the $(2 \to 2)$ rate gradually increases while the $(2 \to 3)$ rate decreases such that the total rate is not sensitive to the scale $Q$. Furthermore, based upon the factorization theorem in the QCD theory, we conclude that the total rates calculated via this prescription will not be sensitive to the choice of PDF although each individual piece will vary with a different set of PDF.

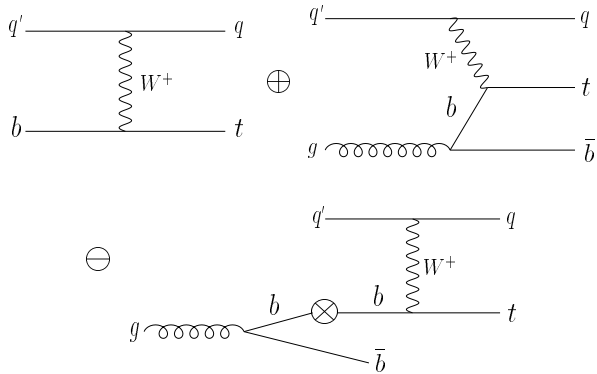

Figure 1: Feynman diagrams illustrating the subtraction procedure for calculating the total rate for $W$-gluon fusion: $q'b \to qt \oplus q'g(W^+g) \to qt\bar{b} \ominus (g \to b\bar{b} \otimes q'b \to qt)$.

---

[b] A complete next-to-leading-order calculation for this process was recently done in Ref. [6] in which the large logs were not properly resummed.

Another single-top quark production mechanism is the Drell-Yan type process $q'\bar{q} \to W^* \to t\bar{b}$. As shown in Table 1, for top quarks with mass on the order of 180 GeV, the rate for $W^*$ production is about one third that of $W$-gluon fusion at $\sqrt{S} = 2$ TeV. The $W^*$ process becomes much less important for a heavier top quark. This is because at higher invariant masses $\sqrt{\hat{s}}$ (for producing a heavier top quark) of the $t\bar{b}$ system, $W^*$ production suffers the usual $1/\hat{s}$ suppression in the constituent cross section. However, in the $W$-gluon fusion process the constituent cross section does not fall off as $1/\hat{s}$ but flattens out asymptotically to $1/M_W^2$. Therefore, for colliders with higher energies, the $W^*$ production mechanism for heavy top quarks becomes much less important. However, the kinematics of the top quarks produced from this process are different from those in the $W$-gluon fusion events [7, 8].

## 3  Measuring the Lifetime of the Top Quark

As shown in Ref. [9], the intrinsic width of the top quark cannot be measured at hadron colliders, such as the Tevatron and the LHC, through reconstructing the invariant mass of the jets from the decay of the top quark produced from the usual QCD processes ($q\bar{q}, gg \to t\bar{t}$) because of the finite resolution of the jet energy measured by the detector. For a 180 GeV SM top quark, its decay width is about 1.6 GeV, however the measured width from the invariant mass distribution of the top quark is unlikely to be better than 10 GeV [10]. Is there a way to measure the top quark width $\Gamma(t \to bW^+)$ to within a factor of 2 or better, at hadron colliders? The answer is yes. It can be measured from the single-top events.

The width $\Gamma(t \to bW^+)$ can be measured by counting the production rate of top quark from the $W$-$b$ fusion process which is *equivalent* to the $W$-gluon fusion process by properly treating the bottom quark and the $W$ boson as partons inside the hadron. In the previous section, we have discussed how to correctly treat the $b$-quark as a parton inside the proton to properly resum all the large logs to all orders in $\alpha_s$. Here, we illustrate how to treat the $W$-boson as a parton inside the proton. Consider the $q'b \to qt$ process. It can be viewed as the production of an on-shell $W$-boson which then rescatters with the $b$-quark to produce the top quark. This factorization is exactly the same as that in the deep-inelastic scattering processes. The analytic expression for the flux ($f_\lambda(x)$) of the incoming $W_\lambda$-boson ($\lambda = 0, +, -$ for longitudinal, right-handed, or left-handed polarization) to rescatter with the $b$-quark can be found in Ref. [11]. The constituent cross section of $ub \to dt$ is given by

$$\hat{\sigma}(ub \to dt) = \sum_{\lambda=0,+,-} f_\lambda\left(x = \frac{m_t^2}{\hat{s}}\right) \left[\frac{16\pi^2 m_t^3}{\hat{s}(m_t^2 - M_W^2)^2}\right] \Gamma(t \to bW_\lambda^+),$$

where $M_W$ is the mass of $W^+$-boson and $\sqrt{\hat{s}}$ is the invariant mass of the hard part process. Since the scattering rate of $Wb \to t$ is proportional to the decay rate of $t \to Wb$, the production rate of single-top event from the $W$-gluon fusion process measures the partial decay width of the top quark $\Gamma(t \to bW^+)$. Furthermore, the branching ratio of $t \to Wb$ can be measured from the ratio of the numbers of double-$b$-tagged versus single-$b$-tagged $t\bar{t}$ events [10]. Combining this model-independent measurement of the branching ratio for $t \to Wb$ with the measurement of the partial decay width $\Gamma(t \to bW^+)$ from the single-top production rate, one can determine the total decay width of the top quark, *i.e.* the lifetime of the top quark. At the Run-II of the Tevatron (with an integrated luminosity of $2\,\text{fb}^{-1}$), we expect that the lifetime of the top quark would be known to about $20\% \sim 30\%$. Here, we have assumed that the branching ratio for $\Gamma(t \to bW^+)$ can be measured to about $10\%$ and the cross section for $W$-gluon fusion process is known to about $15\% \sim 20\%$.

## 4   Probing New Physics via Single-Top Production Rate

The most general operators for the $t$-$b$-$W$ coupling are [12]:

$$\frac{g}{\sqrt{2}}\left[W_\mu^-\bar{b}\gamma^\mu(f_1^L P_- + f_1^R P_+)t - \frac{1}{M_W}\partial_\nu W_\mu^- \bar{b}\sigma^{\mu\nu}(f_2^L P_- + f_2^R P_+)t\right]$$
$$+ \frac{g}{\sqrt{2}}\left[W_\mu^+\bar{t}\gamma^\mu(f_1^{L*} P_- + f_1^{R*} P_+)b - \frac{1}{M_W}\partial_\nu W_\mu^+ \bar{t}\sigma^{\mu\nu}(f_2^{R*} P_- + f_2^{L*} P_+)b\right]$$
$$+ \partial^\mu W_\mu^- \bar{b}(f_3^L P_- + f_3^R P_+)t + \partial^\mu W_\mu^+ \bar{t}(f_3^{R*} P_- + f_3^{L*} P_+)b,$$

where $P_\pm = \frac{1}{2}(1 \pm \gamma_5)$, $i\sigma^{\mu\nu} = -\frac{1}{2}[\gamma^\mu, \gamma^\nu]$ and the superscript $*$ denotes the complex conjugate. In the SM, the only nonvanishing form factor at the tree level is $f_1^L = 1$. These form factors would have different values if new physics exists. Nevertheless, the conclusion that the production rate of the $W$-gluon fusion event is proportional to the decay of $t \to Wb$ holds irrespective of the specific forms of the $f$'s. Hence, measuring the single-top event rate from the $W$-gluon fusion process is an *inclusive* method for detecting effects of new physics which might produce large modifications to the interactions of the top quark. Strictly speaking, from the production rate of single-top events, one measures the sum of all the possible partial decay widths, such as $\Gamma(t \to bW^+) + \Gamma(t \to sW^+) + \Gamma(t \to dW^+) + \cdots$, therefore, this measurement is really measuring the width of $\Gamma(t \to XW^+)$ where $X$ can be more than one particle state as long as it originates from the partons inside the proton (or anti-proton). If new physics strongly enhances the flavor-changing-neutral-current $t$-$c$-$Z$, then the single-top production rate would also be enhanced from the $Z$-$c$ fusion process $qc \to qt$.

Therefore, we conclude that measuring the production rate of the single-top quark event from the $W$-gluon fusion process can probe new physics in an *inclusive* way. This does not require knowing the specific form of the $t$-$b$-$W$ form factors prior to the analysis, in contrast to many other studies (either for the $t\bar{t}$ or the $W^*$ events) which usually require assuming some kind of $t$-$b$-$W$ form factors from the very beginning of the analyses.

## Acknowledgments

This work is supported in part by the NSF under grant no. PHY-9309902.